\documentclass[twocolumn,aps,prl,floats,showpacs]{revtex4}
\newcommand{\beq}{\begin{eqnarray}}
\newcommand{\eeq}{\end{eqnarray}}

\newcommand{\kbar} {\mathchar'26\mkern-9muk}

\usepackage{epsfig}
\usepackage{dcolumn}
\usepackage{bm}

\begin{document}
\bibliographystyle{prsty}
\title{Weak dynamical localization in periodically kicked cold atomic
gases} \draft
\author{C. Tian$^{1}$, A. Kamenev$^{1}$, and A.  Larkin$^{1,2,3}$}

\address{ $^{1}$ Department of Physics, University of Minnesota,
Minneapolis, MN 55455, USA\\
$^{2}$ William I. Fine Theoretical Physics Institute, University of Minnesota,
Minneapolis, MN 55455, USA\\
$^{3}$ L. D. Landau Institute for Theoretical Physics, Moscow, 117940, Russia}

\date{\today}


\begin{abstract}
    {\rm Quantum kicked rotor was recently realized in
    experiments with cold atomic gases and standing optical waves.
    As predicted, it exhibits dynamical localization in the momentum
space.
    Here we consider the weak  localization regime concentrating on
the
    Ehrenfest time scale. The later accounts for the spread-time of a
minimal
    wavepacket and is proportional to the logarithm of the Planck
    constant. We show that the onset of the dynamical localization is
    essentially delayed by four Ehrenfest times and give
    quantitative predictions suitable for an experimental
    verification.
      }
\end{abstract}

 \maketitle



\bigskip

Unprecedented degree of control reached  in experiments with
ultra-cold atomic gases \cite{Chu} allows to investigate various
fundamental quantum phenomena.
A  realization of  quantum kicked rotor (QKR) is one such
possibility that  recently attracted a lot of attention
\cite{Raizen95,Ammann,Raizen99,Zhang04}. To this end cold atoms are placed
in a spatially periodic potential $V_0\cos (2k_Lx)$ created by two
counter-propagated optical beams. The potential is switched on
periodically for a  short time $\tau_p\ll T$, giving a kick to the
atoms; here $T$ is a period of such kicks. The evolution of the
atomic momenta distribution  may be monitored after a certain
number of kicks. If the gas is sufficiently dilute
\cite{footnote1}, one may model it with the single--particle
Hamiltonian, that upon the proper rescaling takes the form
\cite{footnote2} of the QKR:
\begin{equation}
{\hat H}={1\over 2}\,{\hat l}^{\,2}+ K \cos \theta \sum_n \delta
\left(t-n\right)\, .
                                           \label{Hamiltonian}
\end{equation}
Here $\theta \equiv 2k_Lx$ and  time is measured in units of the
kick period, $T$. The momentum operator is defined as $\hat
l=i\kbar\, \partial_\theta$, where the dimensionless Planck
constant is given by $\kbar =8\hbar T k_L^2/(2m)$. Finally, the
classical stochastic parameter is $K = \kbar V_0\tau_p/\hbar \,$.

\begin{figure}
\begin{center}
\leavevmode \epsfxsize=7cm
\epsfbox{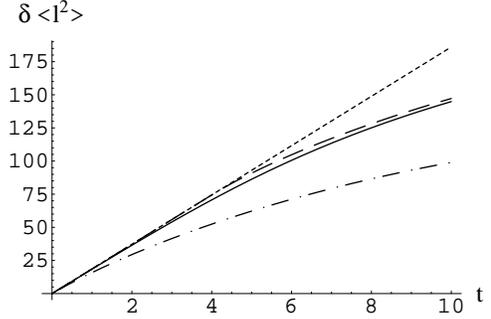}
\end{center}
\caption{ The momentum dispersion for $K=6.1$ and  $\kbar =0.6$ --
full line; the classical limit ($\kbar \rightarrow 0$) -- dashed
line,  standard weak localization ($t_E=0$) -- dashed-dotted line;
the limit $\lambda_2\rightarrow 0$, Eq.~(\ref{result}) --
long-dashed line.} \label{fig1}
\end{figure}

The classical kicked rotor is known to have the  rich and
complicated behavior \cite{Chirikov79}. In particular, for
sufficiently large $K~(\gtrsim 5)$, it exhibits
the chaotic diffusion in the space of angular momentum
\cite{Chirikov79}. The latter is associated with  the diffusive
expansion of an initially sharp  momenta distribution:
$\delta\langle l^2(t)\rangle\equiv \langle
\left(l(t)-l(0)\right)^2\rangle=2D_{cl}t$ (dashed line on
Fig.~\ref{fig1}). For sufficiently large $K$, the classical diffusion constant may be approximated by $K^2/4$ \cite{Chirikov79}.
The higher order correction is an oscillatory function of the stochastic parameter, i.e.,
$D_{cl}(K)\approx {1\over 4}K^2(1-3J_2(K) +2J^2_2(K))$ \cite{Rec81,KFA00}.
It was realized a while ago \cite{CCFI79,CIS81} that quantum
interference destroys the diffusion in the long time limit and
leads to localization: $\delta\langle l^2(t)\rangle\to \sim \xi^2$
at $t\gtrsim t_L\equiv D_{cl}/\kbar^2= \xi^2/D_{cl}$, where the
localization length is given by $\xi = D_{cl}/\kbar$. For a large
localization length $\xi\gg \kbar$, there is a long crossover
regime, $1<t<t_L$, between the classical diffusion and quantum
localization.

It was suggested in Ref.~\cite{FGP82}, that the QKR  may be mapped onto the one--dimensional Anderson localization with the long
range disorder. The universal long--time behavior of the latter is described by the non--linear sigma--model \cite{AZ96},
resulting in the standard weak--localization correction \cite{Altland93,Basko}: $\delta\langle l^2(t)\rangle = 2D_{cl}t\,(1-0.75
\sqrt{t/t_L}\,)$ (dashed--dotted line on Fig.~\ref{fig1}). Notice, that the correction is linear in $\kbar$ and non-analytic in
time. In an apparent contradiction with this fact, explicit studies \cite{She87,DP02} of the first few kicks show only
renormalization of the diffusion constant starting from terms $\sim \kbar^2$. The aim of this paper is to develop a quantitative
description of the classical to quantum crossover for the QKR that, in particular, accounts for these conflicting observations.

It is known in various contexts \cite{LO68} that such crossover
involves an additional time scale, $t_E$, called an Ehrenfest (or
breaking) time. For a generic quantum mapping, it was first shown
by Berman and Zaslavski \cite{BZ78,Zaslavsky81},
that  quantum corrections become comparable to the classical limit
at the time $t_E$. This is the time needed for a minimal quantum
wavepacket, $\delta\theta_0\delta l_0\simeq \kbar$, to spread
uniformly over the angular direction. Due to the chaotic motion,
trajectories diverge as $\delta\theta(t)=\delta\theta_0 e^{\lambda
t}$, where $\lambda$ is the classical Lyapunov exponent. For $K\gg 1$, $\lambda=\ln(K/2)$ \cite{Chirikov79}.
Estimating $\delta l_0\approx K\delta\theta_0$ -- a typical
momenta dispersion after one kick, one finds for the Ehrenfest
time:
\begin{equation}
t_E={1\over \lambda}\ln \sqrt{K\over \kbar}\, .
                                  \label{ehrenfest}
\end{equation}
It is  widely believed \cite{CIS81,BZ78,Zaslavsky81,Izrailev90}
that this intermediate, $1<t_E < t_L$, time scale is indeed
relevant for the quantum evolution of classically chaotic systems.
This observation was put on the quantitative basis  in
Ref.~\cite{AL96} in the context of localization caused by
classical scatterers.  In this paper we adopt methods of
Ref.~\cite{AL96} to the essentially different problem of the QKR.

\begin{figure}
\begin{center}
\leavevmode \epsfxsize=8cm \epsfbox{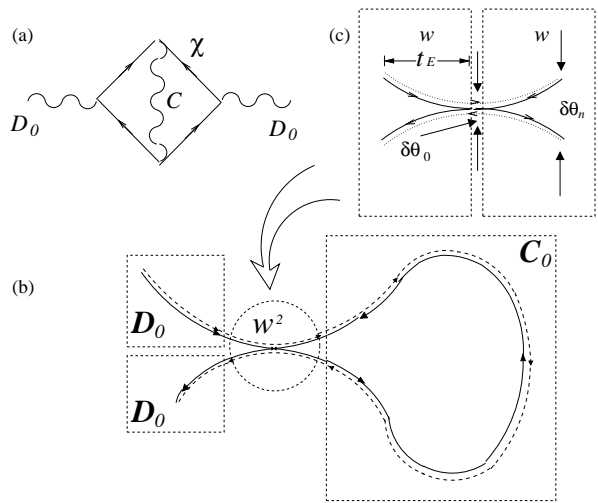}
\end{center}
\caption{ The first quantum correction to the density--density
correlator: (a) one--loop weak localization diagram; (b) its image
in the momentum space; (c) semiclassical Hikami box. }

\label{fig2}
\end{figure}

In the leading order in $\kbar$ we found for the momentum
dispersion:
\begin{equation}
\delta\langle l^2(t)\rangle =2D_{cl}t - \frac{8\kbar
\sqrt{D_{cl}}}{3\sqrt \pi}\,\theta (t- 4t_E) \left(t-
4t_E\right)^{3/2} ,
                                           \label{result}
\end{equation}
where $\theta(t)$ is the step function (long-dashed line on
Fig.~\ref{fig1}). At relatively large time, $t_E \ll t <t_L$ our
result approaches the standard weak--localization, mentioned
above. However, corrections of the order of $\kbar $ are absent
for $t\leq 4t_E$. The delay is caused by the interference nature
of  the localization. Indeed, the first correction originates from
the interference of  two closed--loop counter--propagating
trajectories, see Fig.~\ref{fig2}. It takes time about $t_E$ for
classical trajectories passing through (almost) the same value of
the momentum to diverge and take counter--propagating roots. As a
result, the interference effects are practically absent at smaller
times and show up only after $4t_E$.

One may show that the time interval $0\leq t\leq 4t_E$ is
protected from higher order weak localization corrections as well. For
example, in the second order weak localization correction there are two diagrams \cite{TL03}
proportional to $\kbar^2\, \theta(t-mt_E) (t-mt_E)^2$ with $m=6$
and $m=8$ correspondingly. This fact agrees with the perturbative
studies of the QKR dynamics after a few kicks
\cite{CCFI79,CIS81,Raizen95}, where no localization effects were
seen. (Though the classical diffusion coefficient, $D_{cl}$, is
renormalized as an analytic function of $\kbar^{\,2}$.) It is
important to mention, however, that in reality there is no
non--analyticity at the point $t=4t_E$ as may seem from
Eq.~(\ref{result}). The small localization corrections,
non--analytic in $\kbar$, {\em do} exist for $t \lesssim 4t_E$.
They are associated with the fluctuations of the Ehrenfest time.
Since in the quantum mechanics the minimal separation between the
trajectories is about $\delta\theta_0\sim \sqrt{\kbar/K}$, it
takes a finite time $(\sim t_E)$ for them to diverge. This time
may fluctuate depending on initial conditions. The fluctuations
are characterized by the time scale $\delta t_E =
\lambda_2t_E/\lambda^2$, where (cf. Eq.~(\ref{ehrenfest}))
$\lambda_2=\langle\lambda^2\rangle - \langle
\lambda\rangle^{2}\approx 0.82\,$ for sufficiently large $K$ and the angular brackets denote
averaging over the initial angle. The interference between rare
trajectories, diverging faster than the typical  ones, leads to
quantum corrections at $t\lesssim 4t_E$ of the form:
\begin{equation}
\delta\langle l^2(t)\rangle =2D_{cl}t -\frac{\Gamma
\left(\frac{5}{4}\right)\kbar}{3\pi/256 }  \sqrt{D_{cl}}\, (\delta
t_E)^{3/2} f\left({4t_E-t\over \sqrt{\delta t_E}}\right)\, ,
                                            \label{smalltime}
\end{equation}
where $f(0)=1$ and $f(x)=6\sqrt {2\pi}/ \Gamma
\left(\frac{5}{4}\right) x^{-5/2}e^{-x^2/16}$ for $x\gg1$.
Localization correction including effect of the Ehrenfest time
fluctuations is depicted on Fig.~\ref{fig1} by a full line.  
It is
rather close to the prediction of Eq.~(\ref{result}) (long-dashed
line), however the singularity at $t=4t_E$ is rounded.

Having outlined our main results, we turn now to some details of
the calculations. One starts from introducing the exact one period
evolution operator as: $\hat U=\exp \left\{ i(K/\kbar)
\,\cos\hat\theta\right\}\exp \left\{i\hat l ^2/(2\kbar) \right\}$.
All physical quantities may be expressed  in terms of the matrix
elements of $\hat U^n$, where $n$ stays for the number of kicks
(time). We shall be particularly interested in the four--point
density--density correlator, defined as:
\begin{eqnarray}
&& {\cal D}(l_+,l_-; l_+',l_-';\omega _+,\omega_-)
                                                  \label{QDiff}\\
&\equiv& \sum\limits_{n,n'=0}^\infty \langle l_+|e^{i\omega_+ n}
e^{ \frac{i\hat l ^2}{2\kbar}}{\hat U}^{n}|l_+' \rangle
\overline{\langle l_-'|e^{i\omega_- n'}e^{\frac{i\hat l
^2}{2\kbar}}{\hat U}^{n'}|l_- \rangle}\, ,     \nonumber
\end{eqnarray}
where $|l_{\pm}\rangle$ denote  momentum eigenstates. First we note that averaging over $\left(\omega_++\omega_-\right)/2$ leads
to $n=n'$ \cite{AZ96}. Then performing the standard Wigner transform, one passes to the variables $l=(l_++l_-)/2$ and $\theta$ --
the Fourier transform of $l_+-l_-$ (and similarly for the prime variables). Since the most unstable direction of the underlying
classical dynamics is along the $\theta$ direction \cite{Chirikov79}, any initially smooth distribution quickly relaxes in this
direction. Thus, averaging over $\theta$ and $\theta'$  may be performed. The resulting correlator depends only on the relative
momentum $l-l'$ and frequency $\omega \equiv \omega_+ - \omega_-$. Introducing finally angle $\varphi$ as the Fourier image of
$l-l'$, one ends up with ${\cal D}={\cal D}(\varphi;\omega)$. Its classical limit, ${\cal D}_0(\varphi;\omega)$, may be found e.g.
by using the diagrammatic technique of Ref.~\onlinecite{Altland93}, where in the large $K$ limit it corresponds to the family of
ladder diagrams. Alternatively, one may  show from Eq.~(\ref{QDiff}) that ${\cal D}_0$ satisfies  classical Liouville equation.
Upon the proper regularization \cite{Rec81,KFA00,Zirnbauer99}, that may be viewed as a coarse graining in the angular direction,
one arrives to the classical diffusion propagator:
\begin{equation}
{\cal D}_0 (\varphi;\omega)= \left(-i\omega +D_{cl}
\varphi^2\right)^{-1}\, .
                                                       \label{DF}
\end{equation}
This classical limit reflects the diffusion in the momentum space:
$\delta\langle l^2(t)\rangle=2D_{cl} t$, with the diffusion
coefficient, $D_{cl}(K)$,  studied extensively in the literature
\cite{Chirikov79,Rec81,KFA00}.

The first quantum correction to Eq.~(\ref{DF})  is given by the
one--loop weak--localization diagram, Fig.~\ref{fig2}a.  It
describes the interference of the two counter--propagating
trajectories, passing through (almost) the same point in the
momentum space, Fig.~\ref{fig2}b. In the Wigner representation
such correction takes the form:
\begin{eqnarray}
&&
\delta {\cal D} (l,\theta;l',\theta')=\int\int \frac{dl_0d\theta_0}{2\pi}\,
\frac{dl_1d\theta_1}{2\pi}\,\Big\{
{\cal C} (l_1,\theta _0;l_0,\theta _1)
\,
                                                   \nonumber \\
&\times&
\hat {\cal
X}(l_0,\theta_0;l_1,\theta_1)
\big[ {\cal D}_0(l,\theta;l_0,\theta_0)
\, {\cal
D}_0(l_1,\theta_1;l',\theta')\big] \Big\} \, ,
                                                     \label{qc1}
\end{eqnarray}
where $\omega$ argument is omitted to shorten  notations.
The operator $\hat {\cal X}$ stays for the Hikami box \cite{Hikami81}, which is
given by
\begin{equation}
\hat {\cal X} = -\exp\left\{-\frac{K^2(\delta\theta_0)^4}{4
\kbar^2} +4i\frac{\delta l_0\delta\theta_0}{\kbar}\right\}\,
D_{cl}\left( \nabla^{\,2}_{l_0}+\nabla^{\,2}_{l_1}\right)\, ,
                                                     \label{Hikamibox}
\end{equation}
where $\delta\theta_0\equiv \theta_0+\theta_1$ and $\delta l_0
\equiv \left(l_0-l_1\right)/2$.
 It is clear from this expression that the
quantum correction, Eq.~(\ref{qc1}), is non--zero as long as
$\delta l_0\delta \theta_0\lesssim \kbar\,$ and, therefore, it is
proportional to $\kbar$. The semiclassical Cooperon ${\cal C}
(l_0,\theta _0;l_1,\theta _1)$ gives the probability of return to
(almost) the same momentum, $l_1\approx l_0$, at (almost) the
opposite angle, $\theta_1\approx - \theta_0$. If these conditions
were strict, such motion would be forbidden by the time--reversal
symmetry.
The quantum uncertainty makes it possible. It takes, however, a
long time to magnify the initially small angular variation
$\delta\theta_0\simeq \sqrt{\kbar/K}$ (this estimate as well as
$\delta l_0\simeq \sqrt{\kbar K}$ follows directly from
Eq.~(\ref{Hikamibox})) up to $\delta\theta_n\approx 1$, when the
usual diffusion  takes place.

To take this fact into account we divide the Cooperon trajectory onto two parts: the Ehrenfest region, where $\delta \theta_n \ll
1$ and the diffusive  region with $\delta \theta_n \gtrsim 1$. We denote the corresponding propagators as $\cal W$ and ${\cal
C}_0$ and write in the time representation ${\cal C}(t)=\int dt'\,{\cal W}(t')\,{\cal C}_0(t-2t')$ (cf. Fig.~\ref{fig2}b), where,
as we show below, $t'\approx t_E$. Notice that the diffusive part of the trajectory is shortened by $2t'$, leading to ${\cal
C}(\omega)={\cal W}(2\omega)\, {\cal C}_0(\omega)$.  The diffusive Cooperon, ${\cal C}_0(l_0-l_1;\omega)$, has the same form as
Eq.~(\ref{DF}) and thus ${\cal C}_0(0;\omega)\sim \int d\varphi (D_{cl}\varphi^2-i\omega)^{-1}$.

To evaluate  propagator ${\cal W}(2\omega)$ in the Ehrenfest regime, we define ${\cal W}(z,n)$ as a probability to reach the
deviation $\delta\theta_n\equiv e^z$ during $n$ kicks, starting from an initially small variation, $\delta\theta_0 \simeq
\sqrt{\kbar/K}$. According to the classical equations (the standard map) $\theta_{n}=\theta_{n-1} +l_{n}\,$ and
$l_{n}=l_{n-1}+K\sin\theta_{n-1}\,$, the variation evolves  as $\delta\theta_{n}=\delta\theta_{n-1}(1+K\cos\theta_{n-1})+2\delta
l_{n-1}$.
Since $\delta l_0\simeq K\delta\theta_0 $, in the leading order in $K\gg 1$ the evolution of $\delta\theta$ is given by $
\delta\theta_n \approx \delta\theta_0 \prod_{j=0}^{n-1}K\cos\theta_j$.  Taking the logarithm of this expression, one obtains
\begin{equation}
{\cal W}(z;n) =\! \left\langle \delta\left(z-\ln |\delta
\theta_0|-\sum_{j=0}^{n-1}\ln
\left|K\cos\theta_j\right|\right)\right\rangle ,
                                              \label{W}
\end{equation}
where the averaging is performed over the initial distribution of
$\delta\theta_0$ [with the typical scale $\delta\theta_0\sim\sqrt {\kbar/K}$, cf. Eq.~(\ref{Hikamibox})] as well as over dynamics of the fast variable,
that is $\delta l_n/\delta\theta_n$. For $K\gg 1$, one may treat
$\cos\theta_j$ after successive kicks as independent random
variables and employ the central limiting theorem to perform the
averaging in Eq.~(\ref{W}). As a result,
\begin{equation}
{\cal W}(z;n) \approx \exp\left\{- \frac{(z-\ln \sqrt{\kbar/K} -
n\lambda)^2}{2n\lambda_2}\right\}\, ,
                                              \label{W1}
\end{equation}
where the Lyapunov exponent $\lambda$ \cite{Chirikov79} and its
dispersion $\lambda_2$ are defined as:
\begin{eqnarray}
\lambda &\equiv& \left\langle\, \ln|K\cos\theta| \right\rangle
=\ln (K/2)\, ;
                                           \label{lambda}\\
\lambda _2 &\equiv& \left\langle\, \ln^2|K\cos\theta|
\right\rangle -\lambda^2=\zeta(3)-\ln^2 2\approx 0.82\, ;
                                                 \nonumber
\end{eqnarray}
the angular brackets imply integration over $\theta$. The
Ehrenfest evolution crosses over to the usual diffusion at $\delta
\theta_n\approx 1$, meaning $z\approx 0$. Performing finally the
Fourier transform as ${\cal W}(\omega)\equiv \sum_n e^{i\omega
n}{\cal W}(0,n)$ and employing the definition of the Ehrenfest
time, Eq.~(\ref{ehrenfest}), and the fact that $\lambda_2\ll
\lambda$, one finds:
\begin{equation}
{\cal W}(2\omega)=\exp \left\{2i\omega t_E-\frac{2\omega ^2\lambda
_2 t_E}{\lambda ^2}\right\}\, .
                                                    \label{Gamma}
\end{equation}

Due to the time--reversal invariance there is an exact symmetry
between divergence and convergence of the two classical
trajectories involved in the weak--localization correction. This
symmetry is illustrated on Fig.~\ref{fig2}c. Therefore it takes an
additional time $\sim t_E$ for the two distinct semiclassical
diffusons to arrive to the  point $l_0\approx l_1$ and
$\theta_0\approx - \theta_1$, bringing, thus, another factor
${\cal W}(2\omega)$. In a slightly different language, one may
define  the Hikami box for a classically chaotic system
\cite{AL96,footnote3} as $2\kbar{\cal W}^2(2\omega)
D_{cl}\nabla^2_l$, where one factor ${\cal W}(2\omega)$ comes from
the two legs of the Cooperon, while another originates from the
two diffusons. Finally, the quantum correction, Eq.~(\ref{qc1}),
reduces to the renormalization of the diffusion coefficient in the
classical propagator, Eq.~(\ref{DF}), as $D(\omega) = D_{cl}
+\delta D(\omega)$ with
\begin{equation}
\delta D(\omega) =-{\kbar D_{cl}\over \pi}\, {\cal W}^2(2\omega)\!
 \int \!\! \frac{d \varphi}{-i\omega+D_{cl}\varphi^2}\,\, .
                                                       \label{DDclKW}
\end{equation}
Equations (\ref{Gamma}) and (\ref{DDclKW}) constitute the main
analytical results of this work. They describe quantitatively
dynamical weak--localization of the QKR with the account for the
Ehrenfest time phenomena. One may finally express the time
evolution of the momentum dispersion in terms of the
frequency--dependent diffusion coefficient. The exact relation
reads as:
\begin{equation}
\delta\langle l^2(t) \rangle =\!
\int\limits_{-\infty}^{\infty}\!\! {d\omega\over \pi}\,
\frac{1-e^{-i\omega t}}{\omega^2}\,\,  D(\omega)\, .
                                                 \label{TimeCorrDC}
\end{equation}
Neglecting fluctuations of the Ehrenfest time ($\lambda_2\to 0$ in
Eq.~(\ref{Gamma})) and performing   frequency and angle
integrations in Eqs.~(\ref{TimeCorrDC}), (\ref{DDclKW}),  one
obtains  Eq.~(\ref{result}) for the momenta dispersion. Notice
that in this approximation the evolution is purely classical at
$t\leq 4t_E$. To account for the quantum corrections at $t\lesssim
4t_E$ one needs to keep the $\lambda_2$ term in Eq.~(\ref{Gamma}).
The straightforward integration leads to Eq.~(\ref{smalltime}).

Our results, Eqs.~(\ref{result}), (\ref{smalltime}), are expected
to be quantitatively accurate if  separations between the relevant
time scales: $1<t_E<t_L$ are large enough. This is the case when
the two dimensionless constants satisfy inequalities: $\kbar<1<K$.
(In the experiments, we are aware of \cite{Raizen95,Ammann},
$\kbar\gtrsim 2$, and thus $t_E\lesssim 1$.)
 Another restriction has
to do with the dephasing time, $\tau_\phi$ \cite{Cohen91}. The later may
originate from non--perfect periodicity of the kicks (noise)
\cite{SMOR00}, spontaneous emission \cite{Ammann}, as well as from many--body collisions between the
atoms \cite{footnote1}. Whatever the nature of the
dephasing time, one needs to ensure $t_E<\tau_\phi$ to observe the
weak localization.

We would like to thank A. ~Altland, J. ~Liu and C. ~Zhang for
helpful discussions. Useful comments from S. ~Fishman are gratefully acknowledged.
C.~T. and A.~L. are supported by NSF
Grant No. 0120702; A.~K. is A.~P.~Sloan fellow.


\begin{references}
\bibitem{footnote1}The two--particle scattering  mean
free path is $l_\phi\approx 1/(na^{2})$, where $n$ is the atomic
concentration and $a$ -- the $s$-wave scattering length. The
corresponding dimensionless scattering time is $\tau_\phi =
l_\phi/(Tv)$, where $v$ is a typical atomic velocity, that may be
estimated as $v\approx {\hbar k_L\over \kbar m}|l|\approx {\hbar
k_L\over \kbar m} \sqrt{D_{cl}\tau_\phi}\,$. This leads to the
self--consistent estimate of the dimensionless dephasing time:
$\tau_\phi \approx (l_\phi k_L/K)^{2/3}$.
\bibitem{footnote2}The difference with the conventional QKR,
is in the boundary condition on the wave-function:
$\Psi(\theta+2\pi)=\Psi(\theta)e^{i\theta l_0}$, where $l_0\in
[0,1]$ is a fractional part of an atom's momentum in units $2\hbar
k_L$. It may be eliminated by the gauge transformation that
introduces an Aharonov--Bohm flux  into the kinetic energy $(\hat
l-l_0)^2/2$. Such flux does not affect dynamical localization,
though averaging over $l_0$ helps to eliminate  resonances
\cite{Izrailev90}.
\bibitem{footnote3} In contrast to Ref.~\cite{AL96}, the Hikami box
for the QKR may be found explicitly without  introducing a
regularization.
\end{references}
\end{document}